\documentclass[11pt,twoside]{article}

\usepackage{asp2006}
\usepackage{lscape}

\begin{document}
\title{Identifications and SEDs of the detected sources from the {\sl AKARI} Deep Field South}   %%% Fill in title

\author{Katarzyna Ma{\l}ek$^1$, Agnieszka Pollo$^{2,3}$, Mai Shirahata$^4$, Shuji Matsuura$^{4}$, Mitsunobu Kawada$^5$, and Tsutomu T.\ Takeuchi$^5$}   %%% Fill in author names
\affil{
$^1$ Center for Theoretical Physics, PAS, Poland\\
$^2$ The Andrzej So{\l}tan Institute for Nuclear Studies, Poland\\
$^3$ Jagiellonian University, Krakow, Poland\\
$^4$ Institute of Space and Astronautical Science, JAXA, Japan \\
$^5$ Nagoya University, Japan
}    %%% Fill in author affiliations

\begin{abstract} %%% Abstract to run on from here.
In order to find counterparts of the detected objects in the {\sl AKARI} Deep Field South (ADFS)
in all available wavelengths, we searched public databases (NED, SIMBAD and others).
Checking 500 sources brighter than 0.0482~Jy in the {\sl AKARI} Wide-S band, we found 114 
sources with possible counterparts, among which 78 were known galaxies.
We present these sources as well as our first attempt to construct spectral energy distributions
(SEDs) for the most secure and most interesting sources among them, taking into account all the 
known data together with the {\sl AKARI} measurements in four bands.
\end{abstract}

%%% MAIN BODY OF TEXT GOES HERE. CONSULT "INSTRUCTIONS FOR AUTHORS USING
%%% LATEX2E MARKUP", SECTIONS 2.3-2.6 FOR HELP WITH EQUATIONS, FIGURES,
%%% AND TABLES.

%\section{}   %%% Top level section head (remove "%" symbol)
%\subsection{}   %%% Second level section head (remove "%" symbol)
%\subsubsection{}   %%% Lowest level section head (remove "%" symbol)
%\section*{}    %%% Unnumbered top level section head (remove "%" symbol)
%\subsection*{}   %%% Unnumbered second level section head (remove "%" symbol)

\section{Introduction}   %%% Top level section head (remove "%" symbol)

The {\sl AKARI} Deep Field South (ADFS) is one of the deep fields close to
the Ecliptic Pole.
The unique property of the ADFS is that the cirrus emission density is 
the lowest in the whole sky, i.e., the field is the most ideal sky area for 
far-infrared (FIR) extragalactic observations.
Very deep imaging data were obtained down to $\sim 20$~mJy at $90\;\mu$m
\citep[for details, see ][]{shirahata2009}.

\section{Catalog} \label{sec:catalog}

\begin{figure}[ht]
\begin{minipage}[b]{0.5\linewidth}
   \centering
   \resizebox{0.7\hsize}{!}{
   \includegraphics{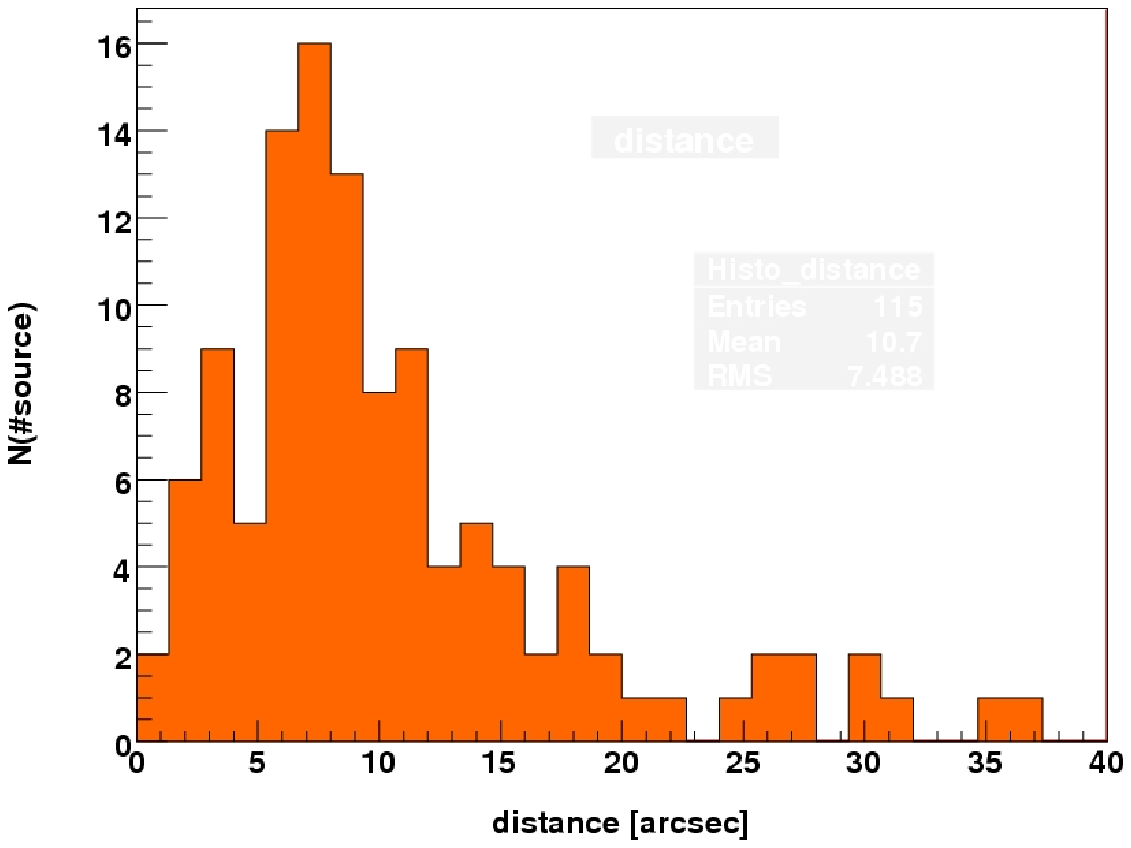}
   }
   \caption{  The distribution of angular deviation of counterparts from the ADFS sources.
   }\label{fig:ddist}
\end{minipage}
%\hspace{0.5cm}
\begin{minipage}[b]{0.5\linewidth}
   \centering
   \resizebox{0.5\hsize}{!}{
      \includegraphics{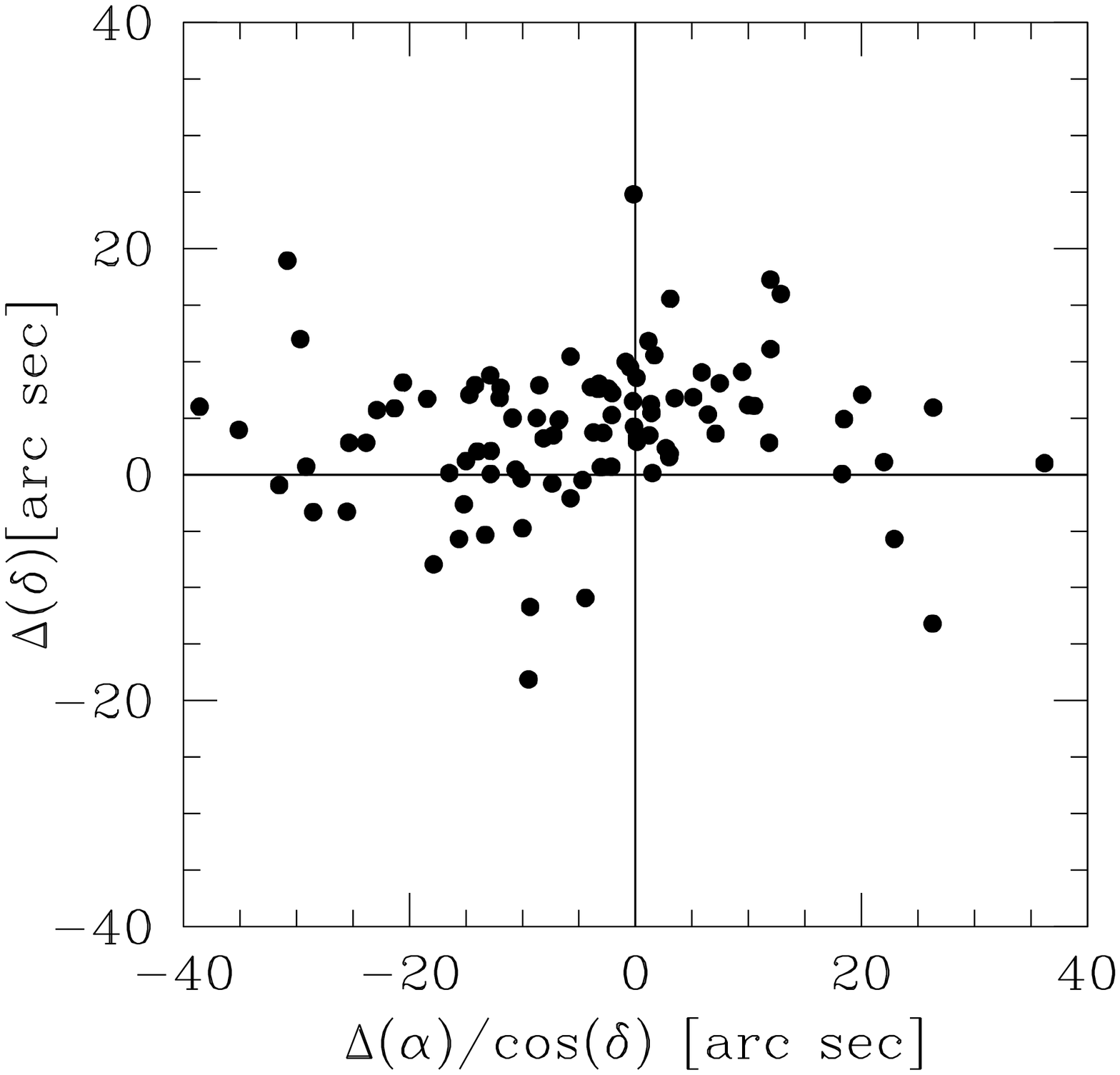}
   }
   \caption{
  The scatter plot of the deviation of 114 counterparts with respect to the ADFS sources.
   }\label{fig:scatter}
\end{minipage}
\end{figure}

We cross-correlated the ADFS point source catalog (based on $90\;\mu$m) with
other known and publicly available databases, mainly the SIMBAD and NED.
For 500 sources brighter than 0.0482~Jy, we searched for their counterparts in other
wavelengths within the radius of $40''$.
In total, 110 counterparts for 114 ADFS sources were found. As shown in Figure~\ref{fig:ddist}, the angular distance between the ADFS source and its counterpart is in most cases smaller than $20''$. It is plausible that the more distant identifications are caused by the contamination. In particular, all the three stars in the sample are most probably falsely identified because of the contamination (M. Fukagawa, private communication). Positional scatter map, shown in Figure~\ref{fig:scatter}, displays a small but systematic bias of $\sim 4 ''$ in declination of the ADFS positions with respect to counterparts.

\begin{figure}[!ht]
\begin{center}
   \resizebox{0.4\hsize}{!}{
     \includegraphics{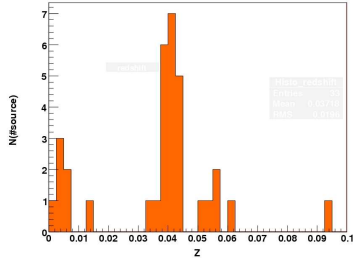}
   }
\end{center}
%\vspace{-5mm}
  \caption{The redshift distribution of 33 identified ADFS galaxies with known redshifts. The only quasar identified (at z=1.053) was not shown here.  
  }\label{fig:zdist}
\end{figure}

We revealed that most of the detected bright sources are galaxies, and very few stars, quasars, or AGNs were found. As shown in Figure~\ref{fig:zdist}, most of the identified objects are nearby galaxies at $z<0.1$, a large part of them belonging to a cluster DC 0428-53 at $z \sim 0.04$. The statistics of identified ADFS sources is presented in Table 1.

%\vspace{-5mm}

\begin{table}[!ht]
\caption{Statistics of identified ADFS sources}
\label{tbl:stat}
\begin{center}
\begin{tabular} {llc}
\hline
\hline
Galaxies & & 78\\
\hline
& Galaxy & 37\\
& Galaxy in cluster of galaxies & 33 \\
%& Cluster of galaxies & 1\\
%& Galaxy in a pair & 1 \\
%& Interacting galaxy & 3 \\
& Pair or interacting galaxies & 4 \\
& Low surface brightness galaxy & 2 \\
& Seyfert 1 & 1 \\
& Starburst & 1 \\
\hline
Star & & 3 \\
\hline
Quasar & & 1 \\
\hline
X-ray source & & 3 \\
\hline
IR sources & & 24 \\
\hline
\end{tabular}
\end{center}
\end{table}

\section{Spectral energy distributions}

Spectral energy distributions (SEDs) give first important clue to the physics
of radiation of the sources.
The deep image at the {\sl AKARI} filter bands has significantly improved
our understanding of the nature of the FIR emission of various sources and 
allows us to update the models of interstellar dust emission.

\begin{figure}[!ht]
\begin{center}
   \resizebox{0.9\hsize}{!}{
     \includegraphics{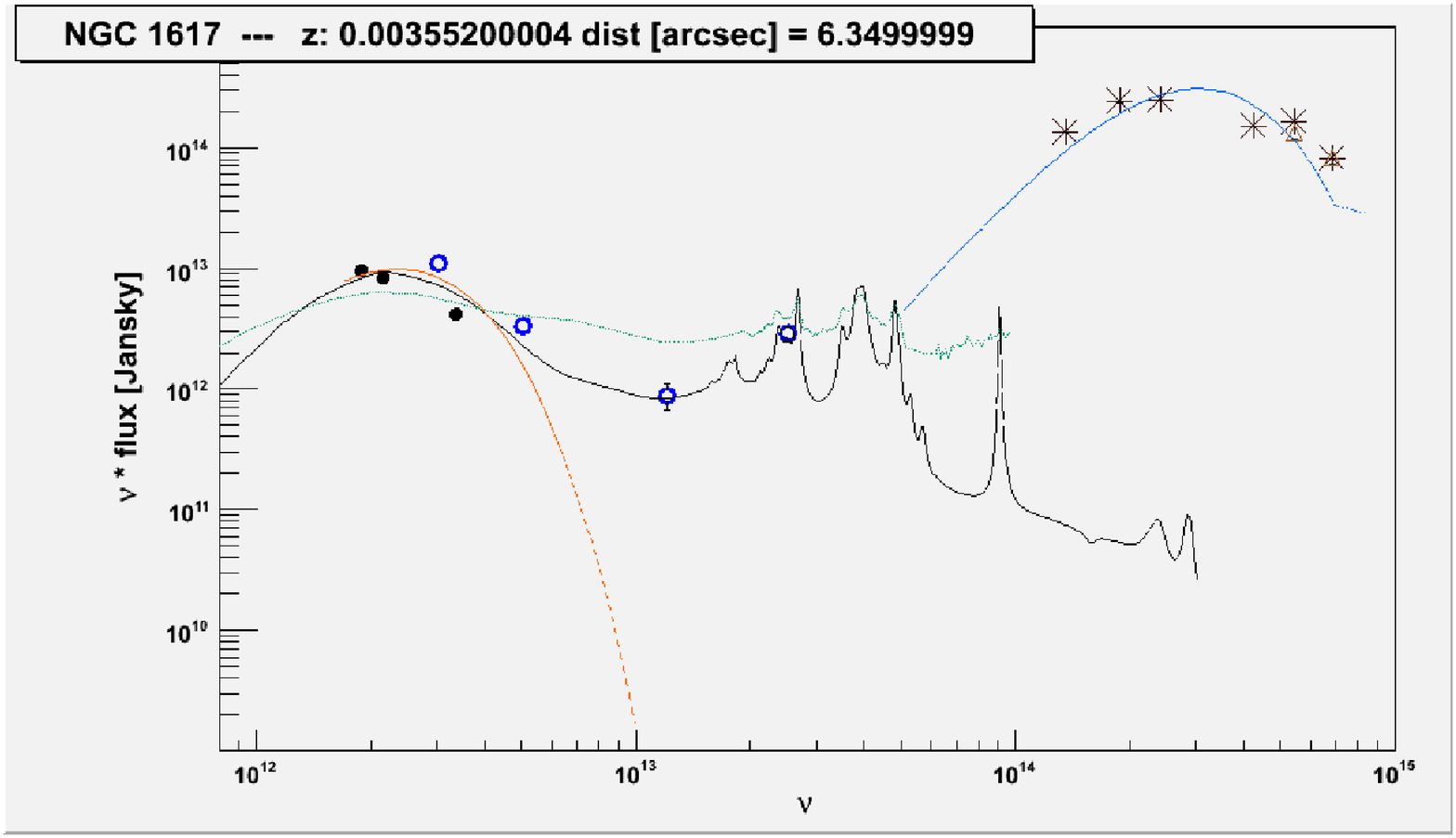}
     \includegraphics{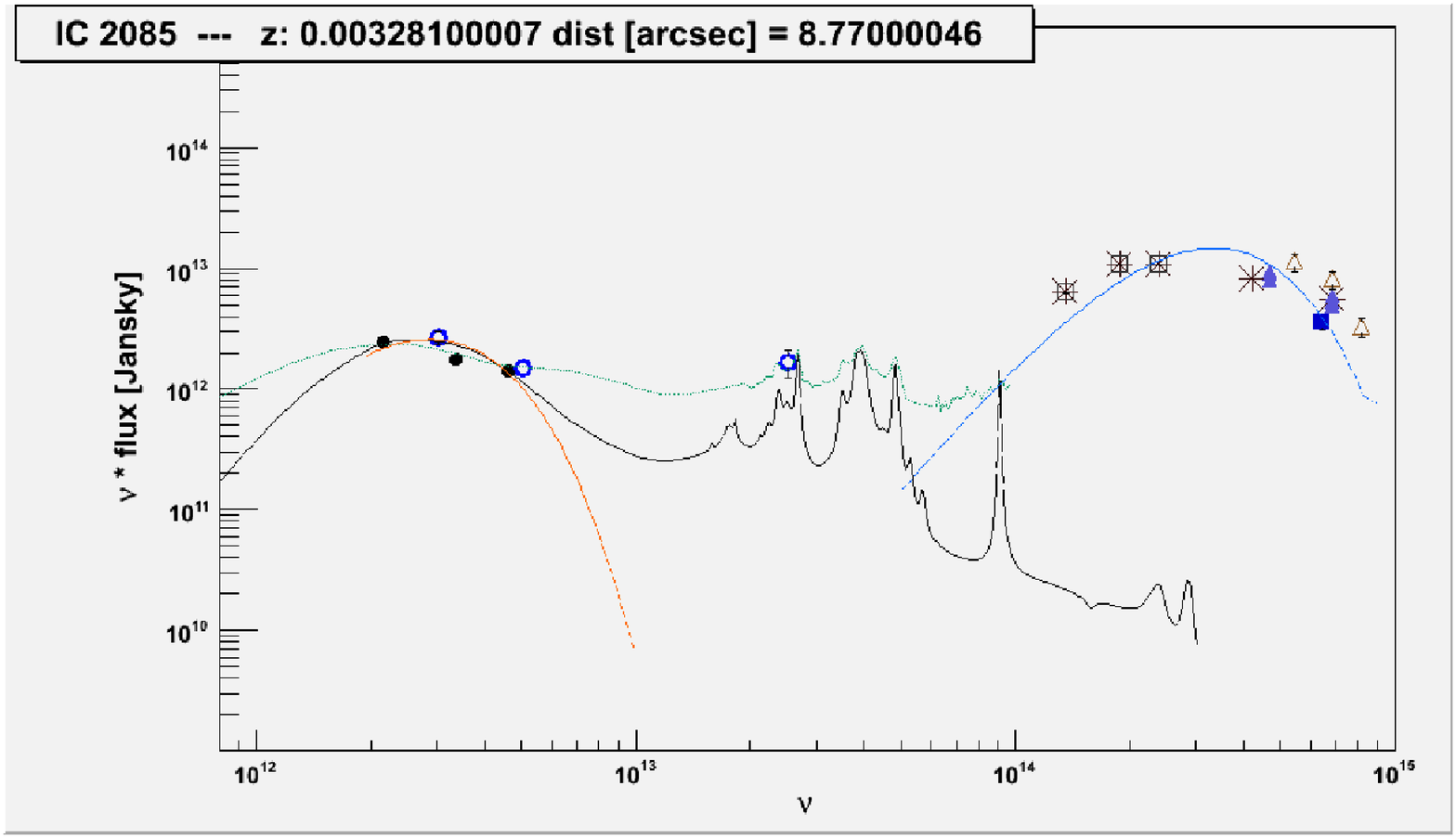}
   }
   \resizebox{0.9\hsize}{!}{
     \includegraphics{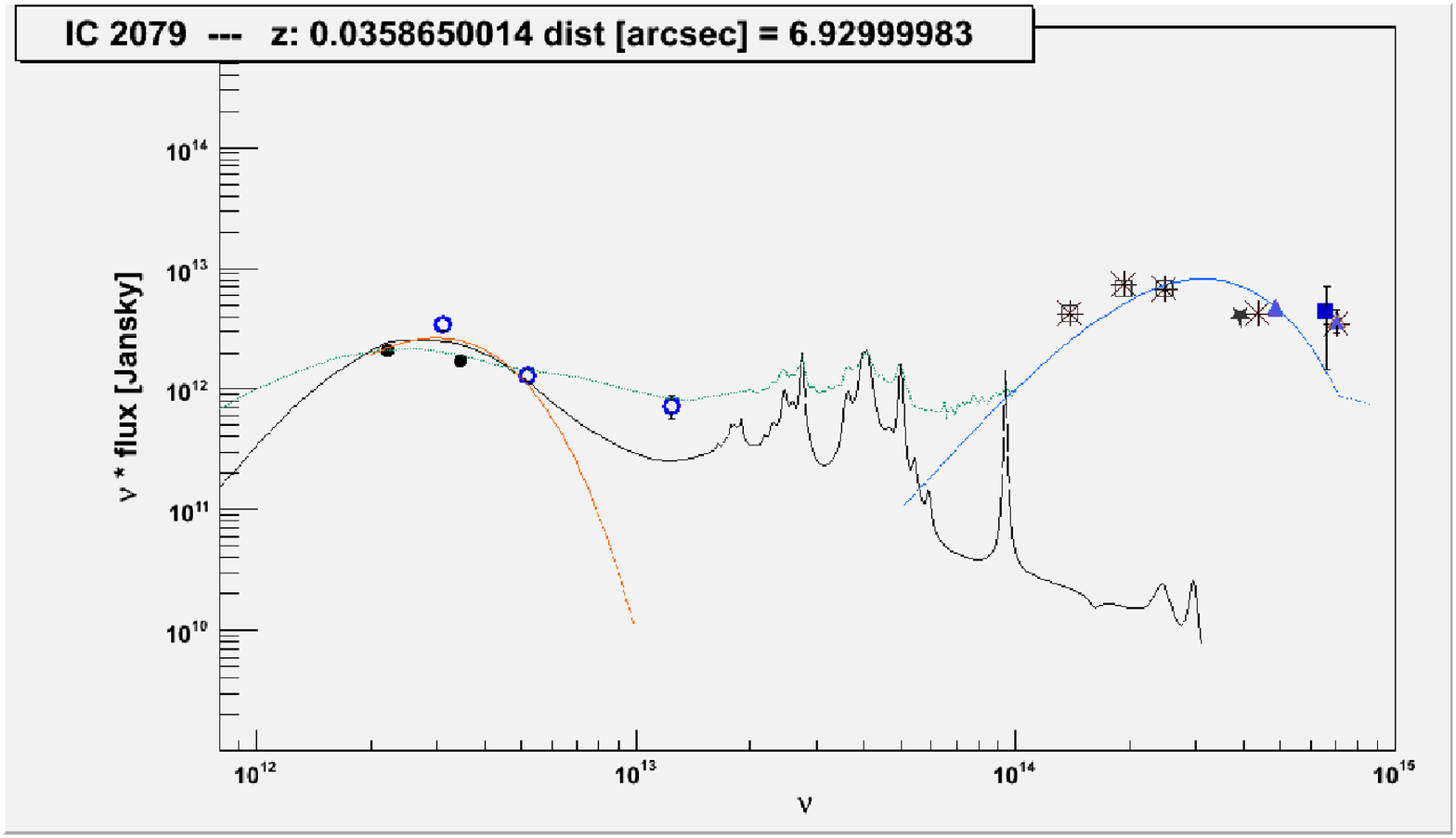}
     \includegraphics{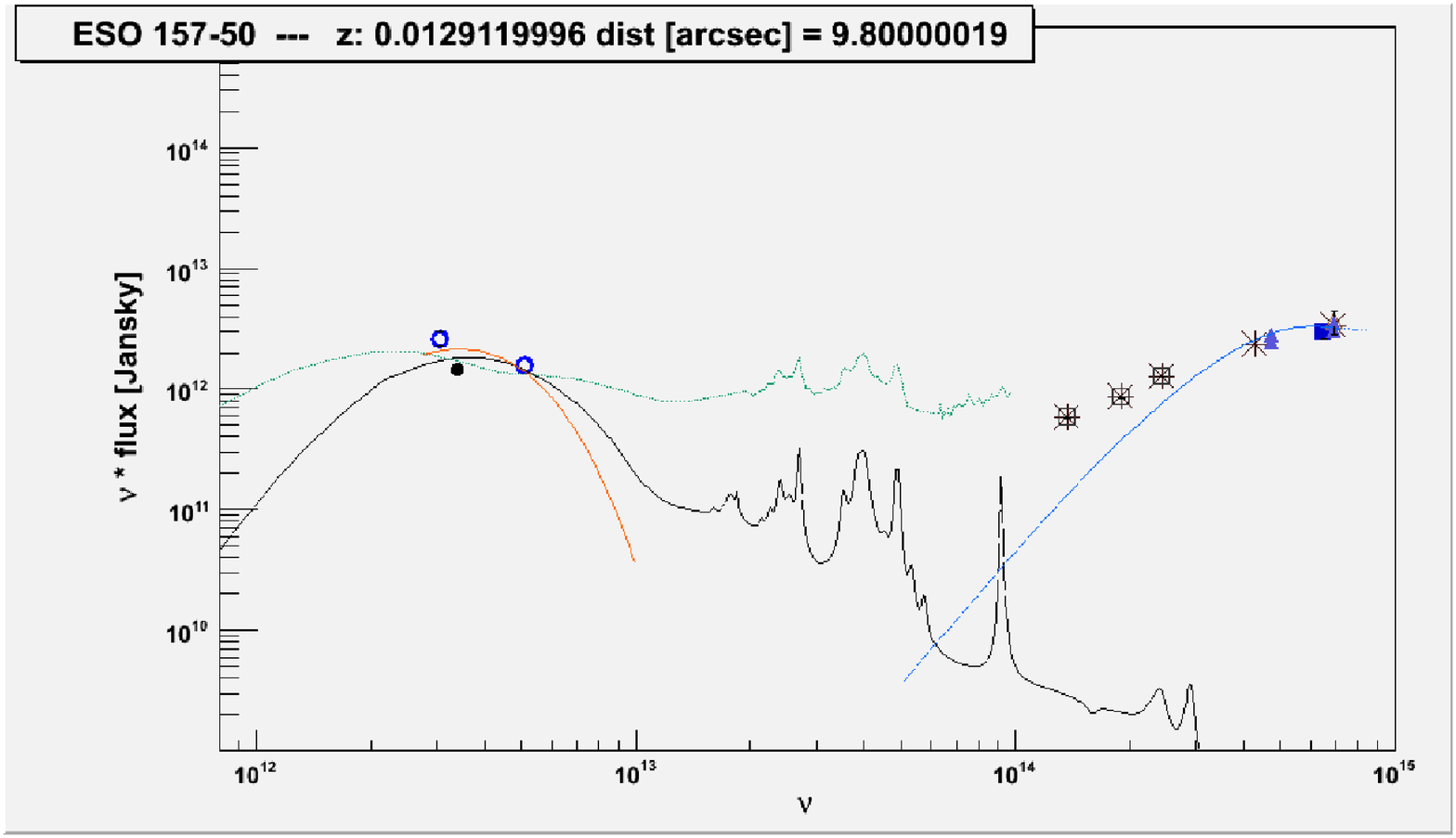}
   }
   \resizebox{0.9\hsize}{!}{
     \includegraphics{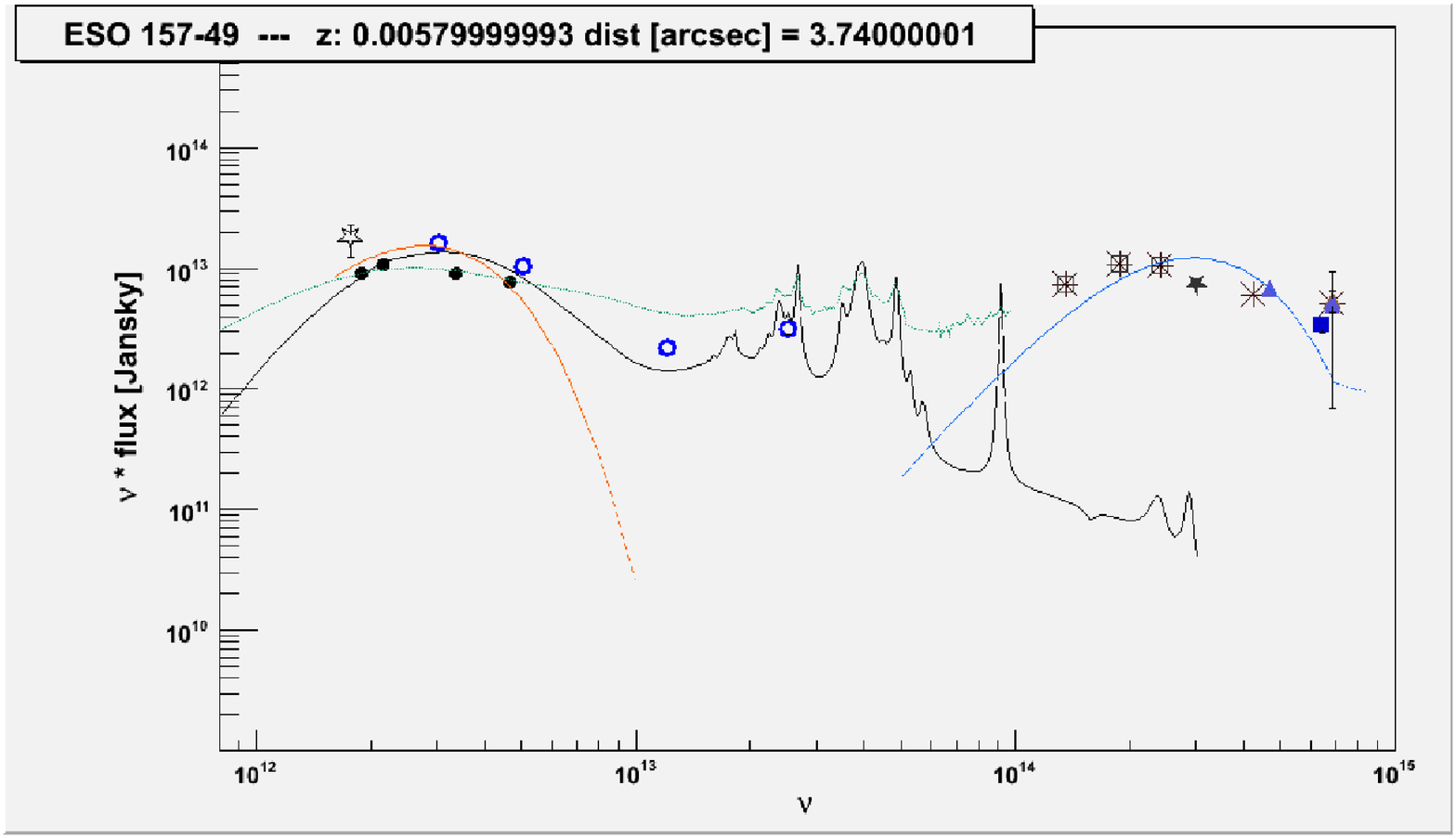}
     \includegraphics{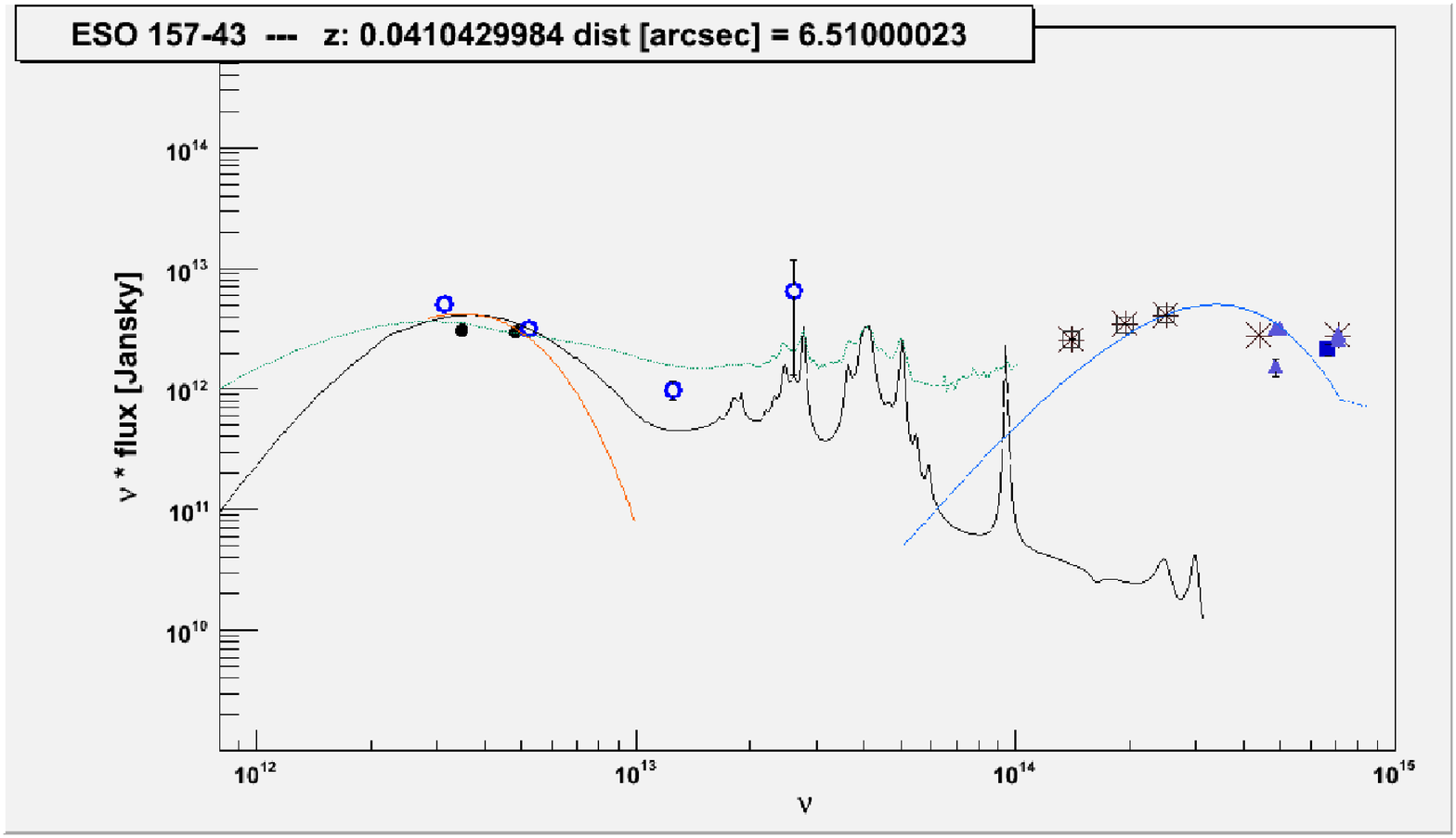}
   }
	
 %  \resizebox{1.0\hsize}{!}{	
%     \includegraphics{kmalek4g.eps}
%     \includegraphics{kmalek4h.eps}
%   }
\end{center}
\caption{
Representative SEDs of galaxies in the ADFS with known redshifts; the data points from ADFS (full black circles) and public databases are fitted by four different models of dust and stellar emission.
}\label{fig:sed}
\end{figure}

We show six representative SEDs of galaxies with known redshifts in 
Figure~\ref{fig:sed}.
We tried to fit four models of dust emission with the SEDs.
First we tried a modified black body ($\nu^\gamma B_\nu(T)$ with 
$\gamma = 1.5$) to the dust emission part, and a black body to the
stellar emission part in the galaxy SEDs ($\nu= 10^{13} \mbox{--} 
10^{14}$~Hz).
Since these galaxies are evolved, a single black body gives a poor
fit to the observed SEDs for some galaxies.
Using a more sophisticated stellar population synthesis model with 
realistic star formation history will be our next step.

For dust emission, we then used models of \citet{dale2002} and
\citet{li2001}.
These more refined models succeeded in reproducing the MIR
part of the dust emission.
By these fittings, we can calculate the mass and temperature of
dust, as well as the PAH contribution to the total dust amount.

\section{NGC 1705}

One of the most interesting objects we found in the ADFS is a nearby
dwarf starburst galaxy NGC 1705.
This galaxy locates at a distance of $5.1\pm 0.6$~Mpc.
It has a relatively low metallicity of $0.35\;Z_\odot$, and the star formation
rate is estimated to be $0.3\;M_\odot \mbox{yr}^{-1}$.
This galaxy is known to harbor the richest super star cluster (SSC) ever
found \citep[][and references therein]{cannon2006}.
The most striking feature of NGC 1705 is that it has completely hidden 
star formation only seen in the FIR, which does not correspond to the 
SSC observed at the optical.
The {\sl AKARI} data, as well as {\sl Spitzer} and {\sl IRAS} measurements
enable more detailed studies on the dust emission, hidden star formation,
ultraviolet radiation field strength and ISM physics of this galaxy.

\begin{figure}[!ht]
\begin{center}
   \resizebox{0.52\hsize}{!}{
     \includegraphics{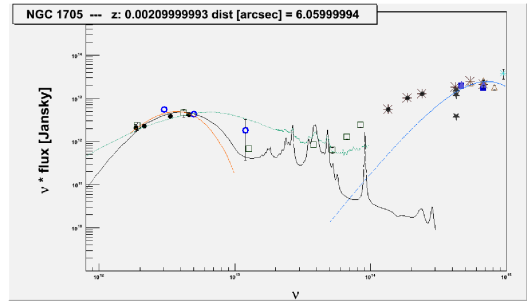}
   }
\end{center}
\caption{
The SED of a dwarf starburst galaxy NGC 1705; the ADFS (full black circles), Spitzer, IRAS and other known measurements are fitted by four different models of dust and stellar emission.
}\label{fig:ngc1705}
\end{figure}

\acknowledgements %%% Text of acknowledgements runs on after this command.

We thank Misato Fukagawa for sending the information about Vega-like 
star candidates. 
This work has been supported (in part) by the Polish Astroparticle Physics Network.
AP was financed by the research grant of the Polish Ministry of Science 
PBZ/MNiSW/07/2006/34A.
TTT has been supported by Program for Improvement of Research 
Environment for Young Researchers from Special Coordination Funds for 
Promoting Science and Technology, and the Grant-in-Aid for the Scientific 
Research Fund (20740105) commissioned by the Ministry of Education, Culture, 
Sports, Science and Technology (MEXT) of Japan.
TTT has been partially supported from the Grand-in-Aid for the Global 
COE Program ``Quest for Fundamental Principles in the Universe: from 
Particles to the Solar System and the Cosmos'' from the MEXT. This research has made use of the NASA/IPAC Extragalactic Database (NED), operated by the Jet Propulsion Laboratory at Caltech, under contract with the NASA and the SIMBAD database, operated at CDS, Strasbourg, France

%%% THE BIBLIOGRAPHY
%%%
%%% CONSULT SECTION 3 OF "INSTRUCTIONS FOR AUTHORS" FOR HOW TO USE NATBIB.
%%% AUTHORS ARE ENCOURAGED TO USE EITHER THE "THEBIBLIOGRAPY" ENVIRONMENT
%%% BY UNCOMMENTING (DELETING THE "%" SYMBOL) THE COMMANDS BELOW, OR BY
%%% USING THE BIBTEX ENVIRONMENT. TO FIND OUT WHICH IS APPLICABLE TO YOUR
%%% CONTRIBUTION, CONSULT THE VOLUME EDITORS FOR YOUR PROCEEDINGS.
%%%


\begin{thebibliography}{}
\bibitem[Cannon et al.(2006)]{cannon2006}
	Cannon, J. M., et al. 2006, \apj, 647, 293
\bibitem[Dale \& Helou(2002)]{dale2002}
	Dale, D. A., \& Helou, G. 2002 \apj, 576, 159
\bibitem[Li \& Draine(2001)]{li2001}
	Li, A., \& Draine, B. T. 2001, \apj, 554, 778
\bibitem[Shirahata et al.(2009)]{shirahata2009}
	Shirahata, M., Matsuura, S., Kawada, M., et al. 2009, this volume

%\bibitem[]{}
%\bibitem[]{}
%\bibitem[]{}
%\bibitem[]{}
%\bibitem[]{}
%\bibitem[]{}
%\bibitem[]{}
\end{thebibliography}
\end{document}